\documentclass[twocolumn,showpacs,preprintnumbers,amsmath,amssymb,prl,aps]{revtex4}

\usepackage{epsfig}% Include figure files
\usepackage{graphicx}% Include figure files
\usepackage{dcolumn}% Align table columns on decimal point
\usepackage{bm}% bold math

\begin{document}
\title{Revelation of the role of impurities and conduction electron density
in the high resolution photoemission study of ferromagnetic
hexaborides}
\author{Kalobaran Maiti}
\altaffiliation{Electronic mail: kbmaiti@tifr.res.in}
\author{V.R.R. Medicherla}
\author{Swapnil Patil}
\author{Ravi Shankar Singh}
\affiliation{Department of Condensed Matter Physics and Materials
Science, Tata Institute of Fundamental Research, Homi Bhabha Road,
Colaba, Mumbai 400005, India}
\date{\today}
\begin{abstract}

We investigate the temperature evolution of the electronic structure
of ferromagnetic CaB$_6$ using ultra-high resolution photoemission
spectroscopy; electronic structure of paramagnetic LaB$_6$ is used
as a reference. High resolution spectra of CaB$_6$ reveal finite
density of states at the Fermi level, $\epsilon_F$ at all the
temperatures and evidence of impurity induced localized features in
the vicinity of $\epsilon_F$, which are absent in the spectra of
LaB$_6$. Analysis of the high resolution spectra suggests that
disorder in B-sublattice inducing partial localization in the mobile
electrons and low electron density at $\epsilon_F$ is important to
achieve ferromagnetism in these systems.

\end{abstract}
\pacs{75.50.Pp, 71.55.Ak, 75.10.Lp, 71.23.-k}
\maketitle

Discovery of unusual ferromagnetism in hexaborides, $M$B$_6$ ($M$ =
Ca, Sr, Ba etc.) \cite{vonlanthen,ott} and in La-doped CaB$_6$
\cite{young} with high Curie temperature (T$_C \geq$ 600~K) has
attracted a great deal of interest due to many interesting
fundamental issues associated to this novel phenomena and huge
potential in technological applications as well. Strikingly, none of
the constituent elements in these compounds possess partially filled
$d$ or $f$ levels to manifest ferromagnetism. In addition, the
ground state of all the ferromagnetic compositions is very close to
the band insulating phase \cite{vonlanthen,ott,young,massida}. One
school believes that ferromagnetism in these systems arises due to
the polarization of low density conduction electrons (electron
density slightly higher than the limit of Wigner crystallization)
\cite{young,ceperley}. Another school describes hexaborides as hole
doped excitonic insulators (excitonic model) \cite{zhitomirsky}. The
later description was prompted by the observation of a band overlap
at $X$ point of the Brillouin zone in {\em ab initio} calculations
\cite{massida} and subsequent description of the de Haas van Alphen
and Shubnikov– de Haas results \cite{goodrich,aronson,hall} within
the same framework. However, more recent {\em ab initio}
calculations \cite{tromp}, and $x$-ray absorption and emission
measurements \cite{denlinger} exhibit insulating phase in CaB$_6$.
Angle resolved photoemission spectroscopic (ARPES) studies indicate
an energy gap $>$~1 eV at $X$ point, which is too high for the
suitability of excitonic model \cite{denlinger,souma}.

In parallel, investigations by Matsubayashi {\em et al.} suggest
that the ferromagnetism in Ca$_{1-x}$La$_x$B$_6$ appears due to Fe
impurities coming from the crucibles used for sample preparation
\cite{matsubayashi}. However, the dependence of ferromagnetism on La
concentration is curious. Subsequently, several experimental and
theoretical studies are carried out to understand the origin of
ferromagnetism in these systems. For example, CaB$_6$ grown using Ca
rich mixture of elements does not exhibit ferromagnetism but becomes
weakly paramagnetic along with an increase in low temperature
resistivity by more than two orders \cite{vonlanthen}. This
observation suggests that ferromagnetism of the nominally pure
system is related to the presence of vacancies in the metal
sublattice. CaB$_6$ grown at different temperatures using high
purity CaO and B show large differences in magnetic and resistivity
behaviors \cite{morikawa}. A recent study \cite{rhyee} shows that
tuning of Ca concentration and/or La doping does not lead to
ferromagnetism if the samples are prepared using high purity
(99.9999\%) boron. Instead, the samples prepared using 99.9\% pure
boron exhibits ferromagnetism at certain compositions. A theoretical
study \cite{monnier} based on {\em ab inito} calculations suggests
that ferromagnetism in Ca$_{1-x}$La$_x$B$_6$ arises due to B$_6$
vacancy and that the dominant contribution comes from the surface
rather than the bulk. Another study attributed ferromagnetism of
CaB$_6$ to $sp$ electrons in narrow impurity bands \cite{edwards}.
Ferromagnetic moment in CaB$_{6}$ is observed to depend strongly on
defect concentration \cite {lofland}.

\begin{figure}
 \vspace{-2ex}
\includegraphics [scale=0.35]{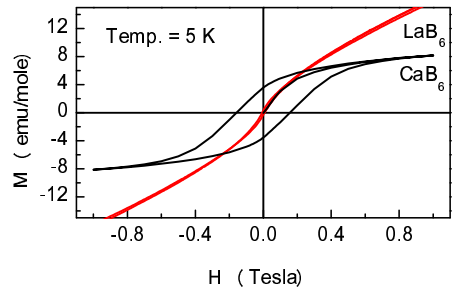}
\vspace{-34ex}
 \caption{(color online) Magnetic hysteresis curve of CaB$_6$ and
LaB$_6$ taken at 5~K.}
 \vspace{-2ex}
\end{figure}

It is clear that ferromagnetism in hexaborides is far from
understood. In this letter, we report our results on the evolution
of the electronic structure of CaB$_6$ and LaB$_6$ in the vicinity
of the Fermi level, $\epsilon_F$ as a function of temperature using
state of the art high resolution photoemission spectroscopy. LaB$_6$
is paramagnetic and hence, is used as a reference. High resolution
spectra of CaB$_6$ exhibit finite spectral intensity at
$\epsilon_F$, despite the fact that it is predicted to be a band
insulator. Temperature evolution reveals impurity features below
$\epsilon_F$ in CaB$_6$, while LaB$_6$ exhibits Fermi liquid
behavior and a disorder induced dip at $\epsilon_F$.

CaB$_6$ and LaB$_6$ were prepared in an arc furnace using 99.7\%
pure B powder, where all the ingredients were kept in a water cooled
copper hearth and melted in ultra high pure argon atmosphere. Thus,
the use of crucibles and hence the source of contaminations could be
avoided. The high quality of the samples were ensured by sharp and
intense x-ray diffraction (XRD) patterns. No trace of impurity was
found in the XRD patterns. Magnetic measurements using high
sensitivity vibrating sample magnetometer (VSM) and SQUID
magnetometer exhibit clear hysteresis loop for CaB$_6$ typical of a
ferromagnetic material as shown in Fig.~1. LaB$_6$, however, does
not exhibit hysteresis loop as evident in the figure. Photoemission
measurements were performed using monochromatized UV source and
electron analyzer, SES2002 from Gammadata Scienta at an energy
resolution of 1.4 meV. The temperature variation down to 10 K was
achieved by an open cycle He cryostat from Advanced Research
Systems, USA. Band structure calculations were carried out using
full potential linearized augmented plane wave (FLAPW) method within
the local density approximations (LDA) using {\scriptsize WIEN2K}
software \cite{wien}. The convergence was achieved considering 1000
$k$ points within the first Brillouin zone. The error bar for the
energy convergence was set to $<$~0.25~meV per formula unit and the
charge convergence achieved was $<$~10$^{-3}$ electronic charge.

\begin{figure}
 \vspace{-2ex}
\includegraphics [scale=0.3]{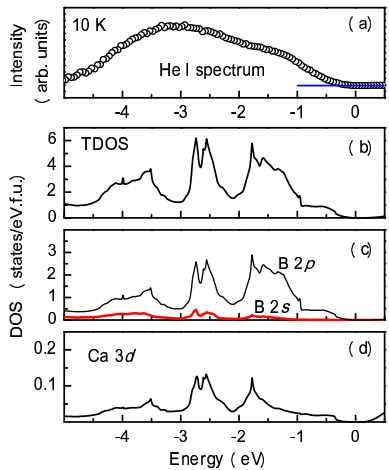}
 \vspace{-8ex}
\caption{(color online) (a) Valence band spectrum of CaB$_6$ at 10
K. Calculated (b) TDOS, (c) B 2$s$ and B 2$p$ PDOS, and (d) Ca 3$d$
PDOS.}
 \vspace{-2ex}
\end{figure}

In Fig.~2(a), we show the valence band spectrum of CaB$_6$ collected
at 10 K using He~{\scriptsize I} radiations. There are two distinct
features around 1.5 eV and 3 eV binding energies as also observed in
previous ARPES study \cite{souma}. Intensity at $\epsilon_F$ appears
to be zero suggesting an insulating character of this material with
a large gap. We show the calculated total density of states (TDOS)
in Fig.~2(b) and various partial density of states (PDOS) in Fig.
2(c) and 2(d). Interestingly, the LDA results exhibit an insulating
ground state with an energy gap of about 0.2 eV characterizing
CaB$_6$ as a {\em band insulator}; the reproduction of the exact
band gap may need other considerations \cite{tromp,hedin}. It is
evident that the TDOS is primarily contributed by B 2$p$ PDOS shown
in Fig.~2(c). All other contributions (B 2$s$, Ca 3$d$, etc.) are
significantly small in this energy range. Similar energy
distribution of Ca 3$d$ PDOS and B 2$p$ PDOS, and the features above
$\epsilon_F$ (not shown here) indicate significant Ca 3$d$-B 2$p$
covalency. The relative intensity of the features in the calculated
results are slightly different from the experimental results
presumably due to the fact that the experimental spectra are
influenced significantly by the matrix element effects, lifetime
broadening of the holes and electrons in addition to various other
final states effects, which are not considered in the {\em ab
initio} calculations. Interestingly, the energy positions of the
features resemble remarkably well with the experimental features.
All these results as well as the observation of similar band
dispersions in previous studies \cite{souma} indicate that the
influence from correlation effects is weak as expected for B 2$p$
electrons.

\begin{figure}
\vspace{-5ex}
 \includegraphics [scale=0.45]{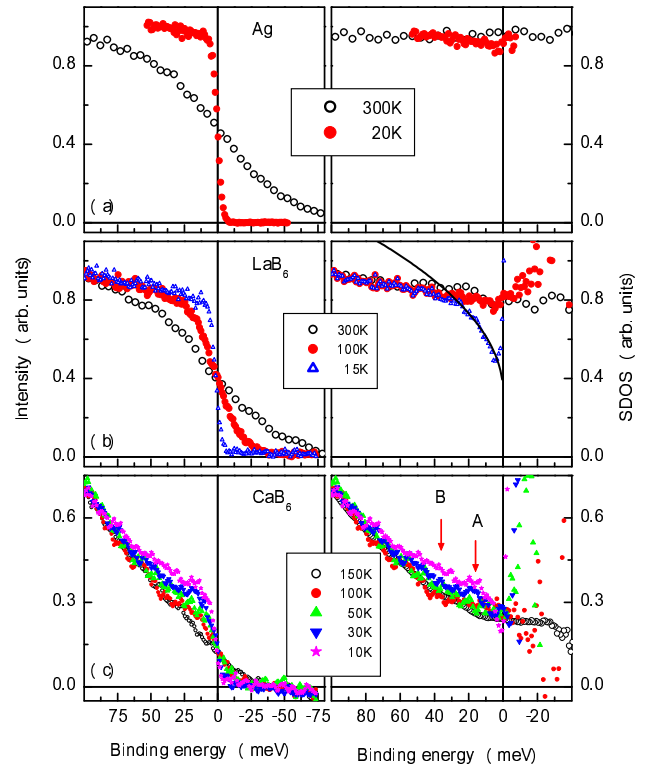}
 \vspace{-8ex}
 \caption{(color online) He {\scriptsize I} spectra of (a) Ag, (b)
LaB$_6$ and (c) CaB$_6$ in the vicinity of $\epsilon_F$. The right
panels exhibit SDOS obtained by dividing the spectra in the left
panel by the Fermi-Dirac distribution function.}
 \vspace{-2ex}
\end{figure}

While the spectrum in Fig.~2(a) exhibits no intensity at
$\epsilon_F$, very high resolution spectra, however, reveal a
different scenario. $\epsilon_F$ at various temperatures was
determined experimentally by the Fermi cut off measured in Ag
mounted on the same sample holder. This is demonstrated in Fig~3(a),
where the valence band spectra of Ag at 300~K and 20~K crosses at
$\epsilon_F$. Interestingly, all the spectra of CaB$_6$ shown in
Fig.~3(c) exhibit finite intensity at $\epsilon_F$, which is
consistent with the observation of Fermi surface in de Haas van
Alphen measurements \cite{hall}. 150 K spectrum of CaB$_6$ exhibits
negative slope near $\epsilon_F$ indicating that  $\epsilon_F$ is
pinned above the valence band. The intensity at $\epsilon_F$ is very
small and presumably arising due to the impurities and/or defects
leading to charge carrier doping in this band insulating material.

All the spectra cross each other at $\epsilon_F$ as expected from
the temperature dependent Fermi-Dirac distribution function.
However, the intensity below $\epsilon_F$ increases unusually with
the decrease in temperature. Such highly anomalous spectral
evolution is not observed in LaB$_6$ as shown in the middle panel of
the figure. We have divided all the spectra in the left panel of
Fig. 3 by the Fermi-Dirac distribution function. The high energy
resolution employed in these measurements introduce negligible
resolution broadening in the experimental spectra. Thus, such
divided spectral functions are a good representation of the spectral
density of states (SDOS) as often observed in other systems
\cite{bairo3,vanadates}. SDOS at different temperatures for all the
three samples, Ag, LaB$_6$ and CaB$_6$ are shown in the right panel
of the figure. SDOS of Ag exhibits a flat distribution of intensity
as a function of binding energy as expected for bulk Ag. LaB$_6$
also exhibits essentially flat and temperature independent SDOS down
to 100 K. Further decrease in temperature leads to a sharp dip at
$\epsilon_F$ with a lineshape, $I(\epsilon) = I_o + a\times|\epsilon
- \epsilon_F|^\alpha$, where $\alpha$ = 0.5 (solid line in the
figure). This suggests that disorder plays an important role in
determining the electronic structure in this system \cite{AA} in
addition to the electron-phonon coupling effect predicted before
\cite{lab6APL,lab6phonon}.

\begin{figure}
 \vspace{-2ex}
 \includegraphics [scale=0.35]{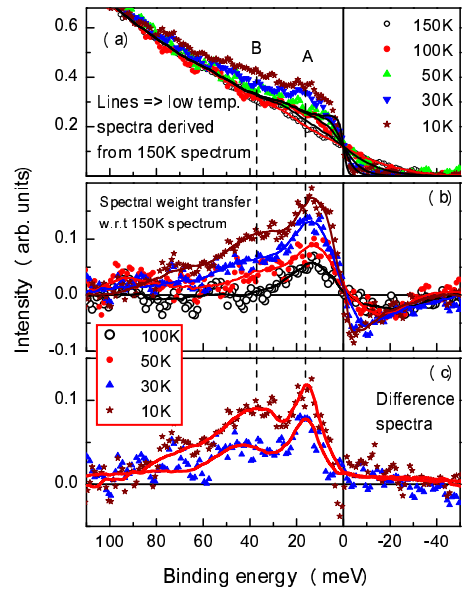}
 \vspace{-4ex}
 \caption{(color online) (a) He {\scriptsize I} spectra of CaB$_6$
at different temperatures. The lines represent the low temperature
spectra derived from SDOS at 150K. (b) The temperature induced
spectral weight transfer calculated by subtracting 150 K spectrum
from all the other spectra. (c) Difference spectra at 30 K and 10 K
obtained by subtracting the lines in (a) from the corresponding raw
data.}
 %\vspace{-4ex}
\end{figure}

Strikingly different behavior is observed in CaB$_6$. The intensity
at $\epsilon_F$ remains almost unchanged down to the lowest
temperature studied. Most interestingly, the intensity below
$\epsilon_F$ increases gradually with the decrease in temperature.
The growing features are manifested by a double peak structure
marked by 'A' and 'B' in the figure. In order to probe the features
more clearly, we have derived the spectral functions at lower
temperatures by convoluting the SDOS at 150 K with Fermi-Dirac
distribution function. The simulated spectra represented by solid
lines are compared with the experimental spectra (symbols) in
Fig.~4(a). Clearly, the experimental spectra exhibit much enhanced
intensity compared to that expected merely due to the thermal
effects included in the Fermi-Dirac distribution function. We have
subtracted the 150 K spectrum from all the spectra at low
temperatures to find the spectral weight transfer, which are shown
in Fig.~4(b). The smooth lines through the points are drawn to guide
the eye. The zero intensity at $\epsilon_F$ indicates that the DOS
at $\epsilon_F$ remain essentially unchanged. Two distinctly
separable features are observed at about 15 meV and 36 meV binding
energies. In Fig.~4(c), we show the difference spectra at 30 K and
10 K after subtracting corresponding estimated spectral functions
shown in Fig.~4(a). Interestingly, the temperature induced changes
at the Fermi edge are not visible in Fig.~4(c). This reveals the
fact that the delocalized density of states follow the Fermi-Dirac
distribution function and may be described within Fermi liquid
picture. The features A and B are weakly localized and manifested at
low temperatures due to less degree of thermal excitations.

These results thus clearly suggest that small impurities
($\sim$~0.3\%) in B-sublattice introduce localized electronic states
just above the valence band. The double peak structure and their
intensity ratio suggest 2$p$ character of the electrons with a
spin-orbit splitting of about 20 meV. This is not unlikely, since
the valence band is constituted essentially by B 2$p$ electronic
states. These features are not observed in LaB$_6$, despite the fact
that both the samples are prepared using identical procedure.
Instead, the spectra in LaB$_6$ exhibit a dip at $\epsilon_F$.
LaB$_6$ is a good metal with large conduction electron density
($\approx$~10$^{22}$ cm$^{-3}$) at $\epsilon_F$, which presumably
smears out the local features suggesting that the conduction
electron density plays an important role here.

The crystal structure of the hexaborides consists of two
interpenetrating cubic lattice formed by metals and B$_6$ octahedra.
The B-sublattice is very stable, robust and essentially B 2$p$
electrons determines the electronic properties. The impurities in B
introduce significant disorder as clearly observed in LaB$_6$ in
addition to doping of charge carriers. This has twofold effects.
While disorder leads to local moments via localization effect, the
delocalized character of the doped carriers helps to mediate
exchange interactions resulting to long range order. This scenario
explains the studies performed so far on various samples. For
example, it is observed \cite{rhyee} that samples prepared with
99.9999\% purity B do not exhibit ferromagnetism despite the tuning
of charge carrier density by changing the Ca-concentration. It is
thus clear that merely change in carrier concentration and/or
vacancies in Ca or B sublattice do not lead to ferromagnetism. On
the other hand, samples such as CaB$_6$, Ca$_{1-\delta}$B$_6$,
Ca$_{1-x}$La$_x$B$_6$ prepared with poorer B purity exhibit
ferromagnetism when the carrier concentration is low. In every case,
the impurities in B-sublattice, vacancies and/or La substitutions
dope low density charge carriers of 2$p$ character. These charge
carriers are partially localized (binding energy $\sim$15 mev) due
to disorder and are responsible for ferromagnetism. It is to note
here that ferromagnetism due to 2$p$ electrons has been predicted in
other systems such as C impurities in BN nanotubes \cite{Wu}.

In summary, high resolution spectra of ferromagnetic CaB$_6$ exhibit
finite density of states at $\epsilon_F$ and emergence of distinct
weakly localized features below $\epsilon_F$, which is not observed
in paramagnetic LaB$_6$. The details of the temperature evolutions
of the high resolution spectra suggests that there are two
parameters that are important to derive ferromagnetism in
hexaborides; disorder in the B sublattice and conduction electron
density.

The authors acknowledge useful discussions with Prof. S. Das Sarma,
University of Maryland, USA and Prof. A. Fujimori, University of
Tokyo, Japan. One of the authors, S.P. thanks the Council of
Scientific and Industrial Research, Government of India for
financial support.

\end{document}